\documentclass[aps,pre,twocolumn,groupedaddress,showpacs]{revtex4}
\usepackage[american]{babel}

\usepackage{subfigure}
\usepackage{graphicx}
\usepackage{amsmath}
\usepackage{amssymb}
\usepackage{amsfonts}
\usepackage[
 bookmarks,bookmarksopen,bookmarksopenlevel=2,bookmarksnumbered=true,
 pdfauthor={Christopher Gaul},pdftitle={Quantum Mechanical Heat Transport in Disordered Harmonic Chains},
 ]{hyperref}				
\usepackage[labelfont=bf,format=hang,justification=centerlast,small]{caption}	
\usepackage[american]{babel}
\addto\captionsamerican{
 }

\newcommand{\sectionname}{chapter}
\newcommand{\subsectionname}{section}
\newcommand{\subsubsectionname}{section}
\def\sSecRef#1{Sec.~\ref{#1}}
\def\ssSecRef#1{Sec.~\ref{#1}}
\def\appendixRef#1{Appendix~\ref{#1}}
\newcommand\subf[1]{\begin{minipage}{0.45cm}
(#1)

 \vspace*{1cm}

\end{minipage}
}

\def\vec{\boldsymbol}
\def\phi{\varphi}
\def\tn{\textnormal}

\def\mean#1{\left\langle #1\right\rangle } 
\def\antikom#1#2{\left\lbrace#1,#2\right\rbrace  }

\def\symmean#1#2{\frac 1 2\mean{\antikom{#1}{#2}}}

\def\menge#1{\left\lbrace#1\right\rbrace  }

\def\arg{\textnormal{arg}}

\def\grad{\vec\nabla}
\def\Re{\tn{Re}}		
\def\Im{\tn{Im}}		

\def\wzwei{.485\linewidth}	
\def\tss{t^{\prime\prime}}	

\def\A{\mathcal{A}}	
\def\B{\mathcal{B}}	
\def\Gth{G_{\textnormal{th}}} 
\def\tt#1{_{\textnormal{#1}}}	
\def\kT{k_B T}		
\def\Oeff#1{{\tilde\Omega_{#1}}}	
\def\oeff{{\tilde\omega}}	
\def\Hc{H\tt{ch}}

\def\Tb{T_{\alpha}}
\def\Tc{T\tt{ch}}

\begin{document}
\title{Quantum mechanical heat transport in disordered harmonic chains}
\author{Christopher Gaul}
\email{christopher.gaul@uni-bayreuth.de}
\author{Helmut B\"uttner}
\affiliation{Physikalisches Institut, Universit\"at Bayreuth, D-95440 Bayreuth, Germany}

\hyphenation{pa-ra-meter}

\begin{abstract}
We investigate the mechanism of heat conduction in ordered and disordered harmonic onedimensional chains within the quantum mechanical Langevin method.
In the case of the disordered chains we find indications for normal heat conduction which means that there is a finite temperature gradient but we cannot clearly decide whether the heat resistance increases linearly with the chain length.
Furthermore, we observe characteristic quantum mechanical features like 
Bose--Einstein statistics of the occupation numbers of the normal modes,
freezing of the heat conductivity and the influence of the entanglement within the chain on the current. For the ordered chain we recover some classical results like a vanishing temperature gradient and a heat flux independent of the length of the chain.
\end{abstract} 

\pacs{05.60.Gg,	
44.10.+i,	
05.10.Gg 	
}

\maketitle

\section{Introduction}
Fourier's law of heat conduction states that the heat flux $j_Q$ through a medium is proportional to the temperature gradient: $\vec j_Q = -\kappa \grad T$. 
In the stationary 1D case, where the heat flux is constant, this implies a constant temperature gradient, at least if the 
thermal conductivity $\kappa$ is not temperature dependent. 

The rigorous deduction of Fourier's law from classical or quantum mechanical statistical physics has been an unsolved problem for decades.
In 1955 Fermi, Pasta and Ulam \cite{Fermi1965} performed the first numerical calculations on the dynamics of harmonic chains, the simplest imaginable model for lattice vibrations in insulators. It was found that harmonic chains do not find their way to equilibrium because the normal modes are decoupled in harmonic systems 
\mbox{[\onlinecite{Rieder1967}\nocite{clLepriUnivers,clLepriLivi03}-\onlinecite{Bonetto2000}]}.
In one dimensional ordered chains the heat flux is independent of the length of the chain and the temperature gradient vanishes. 

In disordered chains the conductivity is reduced due to the Anderson localization of most of the normal modes
\mbox{[\onlinecite{Dyson1953}\nocite{Verheggen1979}-\onlinecite{disoDhar01}]},
leading to a finite temperature gradient. The overall resistance however does not increase linearly with the chain length, but is proportional to its quare root. That means that the specific conductivity still diverges with length.

Non--integrability is necessary for the observation of equilibration of energy and diffusive heat conduction as described by Fourier's law. The problem of heat conductivity in all kinds of model systems is still an active field of research. 
Exemplarily we mention \cite{clSavinGendelman02,clPereiraFalcao06} dealing with nonlinearities and \cite{clBasileBernardinOlla06} dealing with momentum conservation. Nevertheless no model Hamiltonian system, for which Fourier's law could be proven rigorously, has been found yet.

The one dimensional case is easiest to handle and can be partly justified as a model of the homogeneous three dimensional case. However heat diffusion perpendicular to the direction of the heat flux is neglected. This can be compensated with self--consistent heat baths, which add noise and damping with zero average energy flux to every site of the chain. 
\citet*{clBonettoLebowitz04}
find normal heat conductivity in such a model, 
\citet*{Barros2006} study the manipulation of the heat flux in such a system by changing the masses and/or the onsite potentials  in a chain with self--consistent heat baths.

The other side of the problem is the quantum mechanical description of temperature and heat which include thermal fluctuations and dissipation. To achieve this we follow the ansatz by \citet{qbmUllersma66}, for a review see \citet{qbmNieuwen02}. \nocite{qbmHoerhammerBuettner} The considered system is extended by introducing a heat bath consisting of many environment degrees of freedom. These degrees of freedom will be traced out and a Langevin equation is obtained. The energy transfer from the system into the bath degrees of freedom appears as a damping term. Reversely the initial conditions of the bath degrees of freedom appear as a noise term.

The aim of this work is to use the quantum mechanical Langevin ansatz for heat conduction in a chain.
For technical reasons we restrict ourselves to one dimensional harmonic systems without self--consistent heat baths in the ordered as well as in the disordered case. This ansatz was used by \citet{Talkner1990_1,Talkner1990_2} and recently by  \citet{qhcDharRoy06}. Dhar and Roy apply their method to the self--consistent heat bath model \cite{clBonettoLebowitz04} and recover their results in the classical limit.

Alternative systems for the investigation of quantum mechanical heat conduction are systems of coupled spins \cite{mahlerJul05,Saito2003}. The main difference compared to harmonic oscillators is the finite dimensional Hilbert space with e.~g.\ only two energy levels per site for spin--$\frac{1}{2}$. Depending on the types of coupling and of the choice of parameters normal heat conduction is found or not.

\section{Models for heat conduction}\label{sModels}
\subsection{Disordered harmonic chain coupled to two heat baths}\label{ssAllg}
\def\ab{{\alpha}}
We consider a chain consisting of $l$ harmonic oscillators. $X_j$ and $P_j$ denote the coordinate and the momentum of the $j$--th oscillator. The oscillators have common mass $M$ but each oscillator has its own onsite frequency $\omega_j$. Nearest neighbors $X_j$ and $X_{j+1}$ are coupled via the coupling constant $f_j$.
\begin{equation}
\Hc=\sum_{j=1}^l\left( \frac{P_j^2}{2 M}+\frac{1}{2}M\omega_j^2 X_j^2\right) 
+ \sum_{j=1}^{l-1}\frac{f_j}{2}\left(X_j-X_{j+1}\right)^2
\end{equation}
There are two heat baths denoted by $a$ and $b$ which are coupled to the first and to the last oscillator via coupling constants $c_k$.
\begin{align}
H_{a}&=\sum_{k=1}^N\left[ \frac{p_{a k}^2}{2 m_k}+\frac 1 2 m_k \omega_k^2\left( x_{a k}-\frac{c_k}{m_k\omega_k^2}X_1\right)^2 \right]
\\
H_{b}&=\sum_{k=1}^N\left[ \frac{p_{b k}^2}{2 m_k}+\frac 1 2 m_k \omega_k^2\left( x_{b k}-\frac{c_k}{m_k\omega_k^2}X_l\right)^2 \right]
\end{align}
The Hamiltonian of the complete system is then
\begin{equation}
H=\Hc+H_{a}+H_{b}.
\end{equation}
Initially the bath degrees of freedom are independently occupied according to the bath temperatures $T_a$ and $T_b$. The normal coordinates of the isolated chain are occupied according to the temperature $\Tc$. 
Then the couplings $c_i$ are switched on at time $t=0$.

\def\bim{\nu}
\subsubsection{The equations of motion}
Regarding $X_1(t)$ and $X_l(t)$ as known inhomogeneities, the equations of motion for the bath degrees of freedom are solved and substituted into the equations of motion for the chain. One gets the quantum mechanical Langevin equations 
\begin{align}
\ddot X_i(t) =&-C_{ij}X_j(t) + \frac{1}{M}\eta_i(t)\nonumber\\
&-\frac{\gamma(t)}{M} \bim_{i j} X_{j}(0)-\int_0^\infty dt'\frac{\gamma(t-t')}{M}\bim_{i j}\dot X_j(t') ,\label{eqLangevin}
\end{align}
where we use the Einstein notation, i.~e.\ indices appearing twice are implicitly summed over.
The matrix $C$ is the coupling matrix of the isolated chain
$$C_{i j}=\left(\omega_i^2+\frac{f_{i+1}+f_i}{M}\right)\delta_{i j}-\frac{f_{i-1}}{M}\delta_{i-1,j} -\frac{f_{i}}{M}\delta_{i+1,j}, $$
with $f_0$ and $f_l$ set to zero. $\eta_i(t)$ and $\gamma(t)$ are the noise function and the damping function and are discussed below.
The matrix 
$\bim_{i j}=\delta_{i j}\left(\delta_{i 1}+\delta_{i l}\right)$
connects the damping term to the first and the last oscillator of the chain. 

\paragraph{The damping function}
The damping kernel has the form
\begin{equation}\label{gamma}
\gamma(t-t')=\sum_{k=1}^N\frac{c_k^2}{m_k \omega_k^2}\cos(\omega_k(t-t')) .
\end{equation}
If we choose the bath frequencies $\omega_k = k \Delta$, with the level spacing $\Delta$, and the coupling constants $c_k$ according to the Drude--Ullersma--spectrum \cite{qbmNieuwen02}
\begin{equation}\label{drudeullersma}
c_k=\sqrt{\frac{2\gamma m_k\omega_k^2\Delta}{\pi}\,\frac{\Gamma^2}{\omega_k^2+\Gamma^2}}
\end{equation}
and perform the limit $N\rightarrow\infty$ and $\Delta \rightarrow 0$, we get the convenient result
$
\gamma(t)=\gamma\Gamma e^{-\Gamma\left|t\right|}
$
with the Laplace transform
$\hat\gamma(s)=\frac{\gamma\Gamma}{\Gamma+s}$.

\paragraph{The noise functions}
$\eta_a$ and $\eta_b$ 
act on the first, respectively on the last oscillator of the chain
\begin{align}\label{eta}
\eta_i(t)&:=\eta_a(t)\delta_{i1}+\eta_b(t)\delta_{il} .
\end{align}
The noise is determined by the initial conditions of the respective heat bath
\begin{align}
\eta_\ab(t)&:=\sum_{k=1}^N c_k\left[x_{\ab k}(0)\cos(\omega_k t)+\frac{p_{\ab k}(0)}{m_k \omega_k}\sin(\omega_k t)\right] ,
\end{align}
$\ab = a,b$.
In contrast to the damping, the noise functions depend on the temperature of the respective bath.
The random character of the noise function comes from the unknown initial conditions of the bath degrees of freedom.
As the initial conditions $x_i(0)$ and $p_i(0)$ have zero average, the average of $\eta_j(t)$ is also zero. 
Later we will need the symmetrical autocorrelation function of $\eta_\ab(t)$, which is calculated using the initial conditions:
\begin{align}
&K_\ab(t-t'):=\frac{1}{2}\bigl(\mean{\eta_\ab(t)\eta_\ab(t')}+\mean{\eta_\ab(t')\eta_\ab(t)}\bigr) 
\nonumber \\
&=\frac 1 \pi \int_{0}^\infty d\omega \gamma \hbar \omega \frac{\Gamma^2}{\Gamma^2+\omega^2}\coth\left(\frac{\hbar\omega}{2 k_B T_\ab}\right) \cos\bigl(\omega(t-t')\bigr)\label{eqK}
\end{align}
There is no analytical expression for this integral. It is nonzero even for $\Tb=0$, indicating that quantum fluctuation are never absent. In the classical limit $\coth\left(\frac{\hbar\omega}{2 k_B \Tb}\right)$ reduces to ${2 k_B \Tb}/({\hbar\omega})$, and in the Markovian limit $\Gamma \rightarrow \infty$, $K_\ab(t-t')$ becomes $\delta$--like, i.~e.\ white noise. A detailed discussion of quantum noise can be found e.~g.\ in \cite{qbmGardinier91}.

It is worth mentioning that the Heisenberg equations of motion for the operators are identical with the classical equations of motion. All quantum mechanical effects follow from the initial conditions of the chain and the bath degrees of freedom in the noise function.

\subsubsection{The solution of the Langevin equations}\label{ssLsgLangevin}
Thanks to the linearity of the problem the equations of motion \eqref{eqLangevin} can be easily solved in Laplace space, where the convolution turns into a product and the differential equation becomes algebraic. By using the Laplace transform instead of the Fourier transform used in \cite{qhcDharRoy06}, we will be able to solve equation \eqref{eqLangevin} for any $t$ and not only for the stationary case. In Laplace space the equation reads
\begin{align}
s^2 \hat X_i(t) - s X_i(0) - \frac{P_i(0)}{M} =&-C_{i j}\hat X_j  + \frac{\hat \eta_i(s)}{M} \nonumber\\
&- \bim_{i j}\frac{s \hat\gamma(s)}{M}\hat X_j(s) .
\end{align}
We perform a coordinate transformation to the eigenfunctions $Y_i$ of the coupling matrix $C$, i.~e.\ to the normal coordinates of the isolated chain, which are standing waves in the case of the ordered chain. The transformation matrix is denoted by $G$: $X_i=G_{i j}Y_j$, $\left(G C G^t\right)_{i j}=\Omega_i \delta_{i j}$. The equations of motion then read
\begin{align}\label{eqLaplaceYi}
\hat B_{ij}(s) \hat Y_j(s)
&=s Y_i(0)+\frac{Q_i(0)}{M}+\frac{G_{1 i}\hat\eta_a(s)+G_{l i}\hat\eta_b(s)}{M},
\end{align}
with the interaction matrix 
\begin{align}\label{eqInteractionMatrix}
\hat B_{ij}(s):=\left[ (s^2+\Omega_j^2)\delta_{ij}+\frac{s\hat\gamma(s)}{M}\left(G_{1i}G_{1j}+G_{li}G_{lj}\right) \right].
\end{align}

Both damping and noise act on the normal coordinates via $G_{1 i}$ and $G_{l i}$, i.~e.\ their deflections at the ends, where the chain is coupled to the baths.

We need the inverse of the interaction matrix $\hat B(s)$ for solving \eqref{eqLaplaceYi} for $\hat Y_j(s)$. The entries of $\hat B(s)$ are rational functions of $s$.
We extract the common divisor of all matrix entries and end up with a matrix containing only polynomial entries, which can be inverted using Cramer's rule. The entries of $\hat A = \hat B^{-1}$ are rational functions of $s$ with a common denominator. The poles $\lambda_k$ lie in the left half of the complex plane. After performing a partial fraction expansion we transform back to time space, ending up with a sum of decaying exponentials.
In \appendixRef{sRauschAntwort}
we will have a closer look at the inversion of the matrix $\hat B(s)$ in the case of symmetric chains. 

Inverting equation \eqref{eqLaplaceYi} and transforming to time space thus yields


\begin{widetext}
\begin{align}\label{eqYt}	
Y_j(t)&=\sum_{k=1}^l\left[ \dot A_{j k}(t) Y_k(0)+\frac{1}{M}A_{j k}(t)Q_k(0)\right] 
+\frac{1}{M}\int_0^t dt'\left[ F_{j}^a(t-t')\eta_a(t')+ F_{j}^b(t-t')\eta_b(t')\right] ,
\end{align}
\end{widetext}

with the response functions $F_{j}^{a} (t)=\sum_k G_{1 k}A_{j k}(t)$ for the noise $\eta_a(t)$ and $F_{j}^{b} (t) = \sum_k G_{l k} A_{j k}(t)$ for $\eta_b(t)$. The equation for the momenta $Q_j(t)=M\dot Y_j(t)$ is obtained by differentiating.
The response functions are sums of decaying exponential functions, e.~g.\
\begin{align}\label{eqKoeffF}
F_j^\alpha &= \sum_k F_{j,k}^\alpha e^{\lambda_k t} + c.c.\ , &
\Re(\lambda_k)<0 \ , & &\alpha &= a,b .
\end{align}
That means, any contribution from the initial conditions $\vec Y(0)$ and $\vec Q(0)$ vanishes with time. As mentioned above, the averages $\mean{\eta_a(t)}$ and $\mean{\eta_b(t)}$ are zero, so $\mean{Y_j(t)}$ is zero, as well. Two--point--correlations of coordinates and momenta are the objects of interest.

\subsubsection{The evaluation of time dependent correlations}\label{2erKetteKorrelationen}
With the response functions $A_{j k}(t)$ and $F_j^\alpha(t)$, and the symmetrical autocorrelation functions $K_\alpha(t-t')$ of the noise provided, one can evaluate any symmetrical correlation like $\symmean{Y_i(t)}{Y_j(t)}$ or $\symmean{Y_i(t)}{Q_j(t)}$. E.~g.
\begin{widetext}
\begin{align}
\symmean{Y_i(t)}{Y_j(t)}&=\sum_{k,n}\biggl[
	    \dot A_{ik}(t) \dot A_{jn}(t) \mean{Y_k(0)Y_n(0)} 
+\frac{1}{M^2}   A_{ik}(t)	A_{jn}(t) \mean{Q_k(0)Q_n(0)}\nonumber\\
&\qquad\qquad\qquad+\frac{1}{M}
      \left(\dot A_{ik}(t) 	A_{jn}(t)
	   +     A_{ik}(t) \dot	A_{jn}(t)\right)\symmean{Y_n(0)}{Q_k(0)}
\biggr] \nonumber\\
&\quad +\frac{1}{M^2}\int_0^t dt'\int_0^{t} d\tss \biggl[ 
F_{i}^a(t-t')F_{j}^a(t-\tss)K_a(t'-\tss)+
F_{i}^b(t-t')F_{j}^b(t-\tss)K_b(t'-\tss)
\biggr]\label{eqAllgYYKorrelation}
\end{align}
\end{widetext}
The generalization to time--shifted correlations is cumbersome, but straight forward.

In the integral there is a summation over the exponentials $\exp(\lambda_k(t-t'))$ and $\exp(\lambda_{k'}(t-\tss))$ from \eqref{eqKoeffF} 
and the $\omega$--integration in $K(t'-\tss)$, see \eqref{eqK}. The double time integration over exponentials and cosines can be performed analytically. In the limit $t\rightarrow\infty$ it yields the 
result 
$$
\frac{\lambda_k \lambda_{k'} + \omega^2}{(\lambda_k^2+\omega^2)(\lambda_{k'}^2+\omega^2)}\ .
$$
Then the $\omega$--intergations are evaluated for each pair of roots ($\lambda_k,\lambda_{k'}$) and summed up.

Beside calculating the roots $\lambda_k$ and the coefficients for the response functions $F_{j,k}^\alpha$ \eqref{eqKoeffF}, the evaluation of the $\omega$--integrals is the main numerical work. For both tasks we have employed standard routines from the \textit{Mathematica} environment

\subsection{Symmetric disordered chains}\label{ssSymm}
In this \subsectionname\ we consider a specialization of the system above, namely a disordered chain with left--right symmetry, which means that the Hamiltonian is invariant under the exchange $X_n \rightarrow X_{l+1-n}$. This implies $f_n=f_{l-n}$ and $\omega_n = \omega_{l+1-n}$. The normal coordinates of $\Hc$ are either symmetric or antisymmetric with respect to commuting left and right. In particular the transformation matrix $G$ and the noise response functions obey the following relations
\begin{equation}\label{eqSymmG}
\begin{aligned}
Y_i &\text{ even }& &\Rightarrow & G_{1i}&=\ \ G_{li} & &\Rightarrow & F_i^a(t) &=\ \ F_i^b(t)\\
Y_j &\text{ odd } & &\Rightarrow & G_{1j}&=   -G_{lj} & &\Rightarrow & F_j^a(t) &=   -F_j^b(t)\ .
\end{aligned}
\end{equation}
Even modes thus respond to the effective noise $\eta_e=\eta_a+\eta_b$ and odd ones to $\eta_o=\eta_a-\eta_b$. 
The nondiagonal terms in the interaction matrix $\hat B(s)$, see \eqref{eqInteractionMatrix}, are proportional to $\left(G_{1i}G_{1j}+G_{l i}G_{l j}\right)$ and vanish, if the corresponding modes have different symmetry.
Thus $\hat B_{i j}(s)$ and its inverse $A$ are block matrices which do not mix even and odd modes, allowing to invert $B$ separately for each symmetry family with separate sets of roots.

\def\p#1{p({#1})}
\def\sf#1{\sigma({#1})}
For convenient notation we define the symmetry function $\p{j}$ and the symmetry family $\sf{j}$
\begin{align}
\p{j}&:=
\begin{cases}
e \quad \text{for } Y_j \text{ even} \\
o \quad \text{for } Y_j \text{ odd}
\end{cases} &
\sf{j}&:=\menge{i\ |\ \p{i}=\p{j}}\, .
\end{align}

With $\hat F_{j}(s):=\hat F_j^a(s)$
one can express equation \eqref{eqYt} in the following compact form
\begin{align}\label{eq:allgKetteGetrenntesRauschen}
Y_j(t)&=\sum_{k\in\sf{j}}\left[ \dot A_{jk}(t) Y_k(0)+\frac{1}{M}A_{jk}(t)Q_k(0)\right] \nonumber\\
&\quad +\frac{1}{M}\int_0^t dt'F_{j}(t-t')\eta_{\p{j}}(t')\ .
\end{align}

The noise response functions have the same form as in \eqref{eqKoeffF}, but
only exponentials with $\lambda_k$ from the same symmetry as $Y_j$ occur.

\subsection{Ordered Chains}
A further specialization of the disordered chain is to set all couplings $f_n$ equal to a constant $f$ and all $\omega_n$ to $\omega_0$. In this case the normal coordinates of the isolated chain are simply standing waves with anti--nodes at the ends.

\section{Results}
In this \sectionname\ we present and discuss some numerical results obtained with the techniques presented in the previous \sectionname.

At first we will investigate chains without disorder, i.~e. $f_i = f$ and $\omega_i=\omega_0$. In \sSecRef{ssResultsDisorder} we will introduce disorder in the couplings $f_i$.
The onsite frequencies $\omega_i$ are kept ordered. They cannot be set to zero because the translation of the center--of--mass must be suppressed. We set all $\omega_i$ to a constant $\omega_0$, meaning that our chain is fixed on a substrate and cannot move macroscopically. Another possibility, often employed in the literature \cite{Talkner1990_1,Talkner1990_2,clLepriLivi03}, is to fix only the first and the last oscillator with an onsite potential, corresponding to a free wire spanned between the heat baths. At some points we will refer to this model as well.

In the numerics we work with dimensionless quantities. The mass $M$ and the onsite frequency $\omega_0$ fix together with the Planck constant $\hbar$ and the Boltzmann constant $k_B$ the units of all quantities. Thus, in the results frequencies are given in units of $\omega_0$, energies in $\left[E\right]=\hbar\omega_0$, temperatures in $\left[T\right] = \hbar\omega_0/k_B$, currents in $\left[J\right]=\hbar\omega_0^2$, conductivities in $\left[G\right]=\omega_0k_B$,
coupling constants in $\left[f\right]=M\omega_0^2$, 
momenta in $\left[ P \right] = \sqrt{M\hbar\omega_0}$ and lengths in 
$\left[X\right] = \sqrt{\hbar/(m \omega_0)}$.

Furthermore we fix the cutoff $\Gamma=10\omega_0$, so the couplings $f$ and $\gamma$ and the temperatures are the free parameters. Unless specified otherwise we choose a chain with length $l = 20$.

Except for \sSecRef{ssTime}, we will focus on the correlations within the coordinates and momenta of the chain in the stationary regime, i.~e.\ the initial conditions of the chain are irrelevant.

\subsection{Ordered Chains}\label{ssResultsOrder}
In contrast to 
\sectionname\ \ref{sModels}, where we started with the general disordered case and specialized the system untill the ordered chain, we will begin with the simplest case, the ordered chain, here.

\subsubsection{Energy distribution in the normal coordinates}
As the calculation of the variances is performed in normal coordinates, we inspect the energy distribution in normal coordinates first. 
The energy in the normal coordinates of the unperturbed chain reads $E_{i}=\frac{1}{2}M\Omega_i^2 \mean{Y_i^2} +\frac{1}{2 M}\mean{Q_i^2}$. As the coupling to the heat baths is rather strong, we take the coupling energy into account by determining effective frequencies.
Therefore we perform a second calculation with both bath temperatures set to zero, i.~e.\ the entire system will be in the ground state. We then identify the groundstate energy with the zero point energy of an oscillator with the effective frequency $\Oeff{i}$.
Furthermore we assume that the virial theorem is approximately valid, although the normal coordinates of the unperturbed chain are coupled among each other by the interaction with the baths. Therefore one can calculate effective frequencies $\Oeff{i}$, using only the kinetic energies $\mean{Q_i^2}/(2M)$:
$$
(E_0)_i \approx 2 (E^\text{kin}_0)_i =\frac{\mean{Q_i^2}_0}{M}= \frac{1}{2}\hbar \Oeff{i}
$$
Then we use the effective frequencies $\Oeff{i}$ to calculate the occupation numbers from the kinetic energies at finite temperatures:
$$
n_i = \frac{\mean{Q_i^2}_0/M}{\hbar\Oeff{i}}-\frac{1}{2}
$$
The low frequency modes with large amplitudes at the ends of the chain are affected the most by the coupling to the baths.
\begin{figure}[bthp]
\subfigure[]{\includegraphics[width=\wzwei]{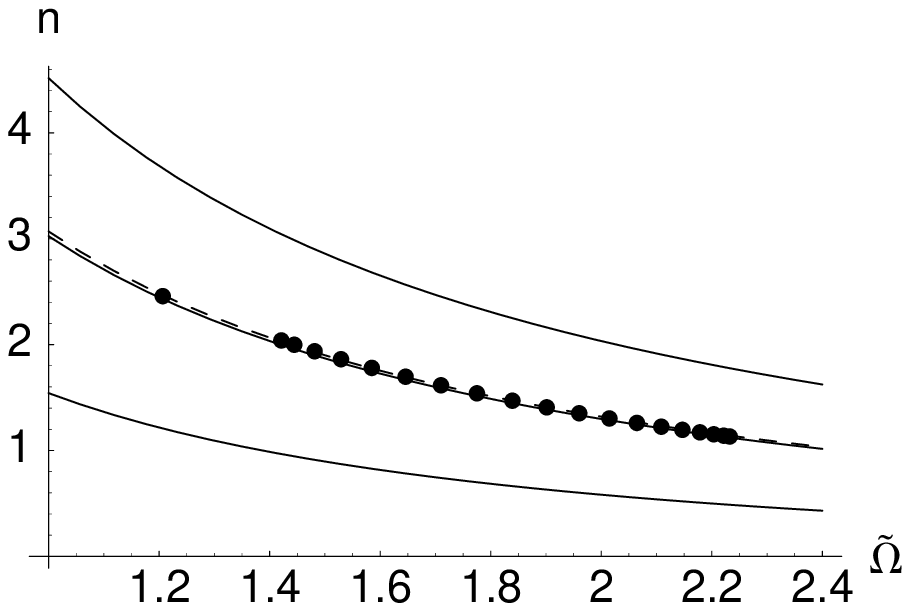}} \hfill
\subfigure[]{\includegraphics[width=\wzwei]{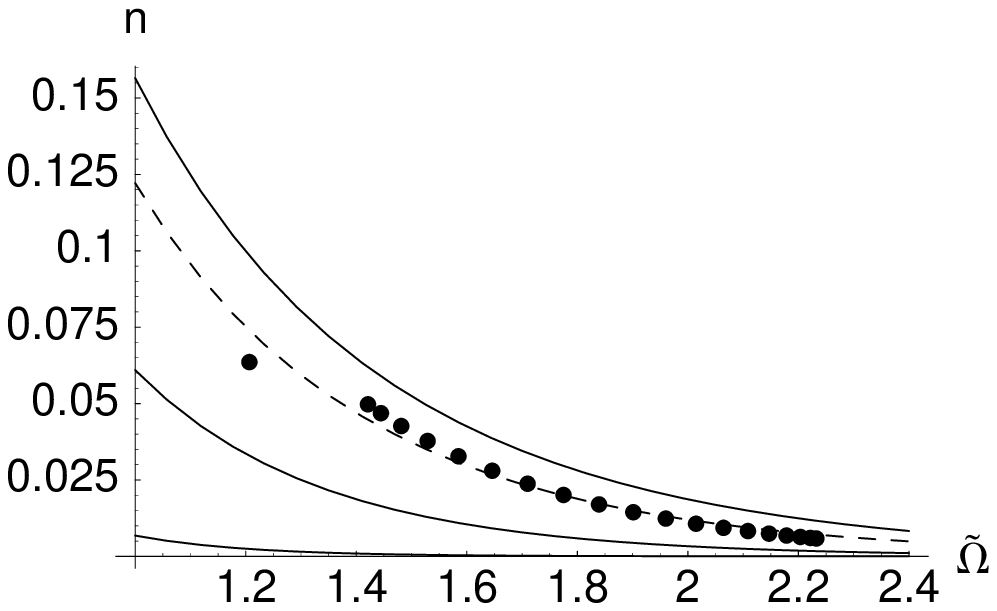}}
\caption{Occupation numbers of the normal modes calculated from the kinetic energies. Parameters: $f=1$, $\gamma= 2$. 
(a) high temperatures $T_a=5$, $T_b=2$, $T\tt{fit} =3.54513$
(b) low temperatures $T_a=0.5$, $T_b=0.2$, $T\tt{fit} = 0.450906$. The center--of--mass motion shows a negative deviation.}\label{figOccNumbers}
\end{figure}

In \autoref{figOccNumbers} these occupation numbers are plotted. Although there is a temperature difference and a heat flux between left and right (see \ssSecRef{ssHeatFlux}), the occupation numbers essentially agree with the Bose--Einstein distributions $\left[ \exp\left(\frac{\hbar \Oeff{}}{k_B T\tt{fit}}\right)-1\right]^{-1}$ in both cases. 
In the high temperature case (a), $T\tt{fit}$ agrees with the average temperature $(T_a+T_b)/2$.
In the low temperature case (b), the temperature $T\tt{fit}$ is closer to the higher bath temperature, giving a hint to the fact that the heat conductivity increases with temperature (\sSecRef{ssHeatFlux}). There are some deviations from the Bose--Einstein distribution resulting from the approximations made in the calculation of the frequencies. In the limit of weak coupling they vanish.

Different definitions of effective frequencies are possible. 
For example one could use the potential energies instead of the kinetic energies and would obtain different results. 
The reason for this is that the normal coordinates of the unperturbed system 
are coupled via the heat baths. We are not dealing with independent oscillators in the ground state, which is reconfirmed by the fact that we find $\mean{Y_i^2}_0\mean{Q_i^2}_0 > \left( \frac{\hbar}{2}\right)^2$.

\subsubsection{Temperature profiles}\label{ssTempProfiles}
\begin{figure}[bthp]
{\subfigure[]{\includegraphics[width=\wzwei]{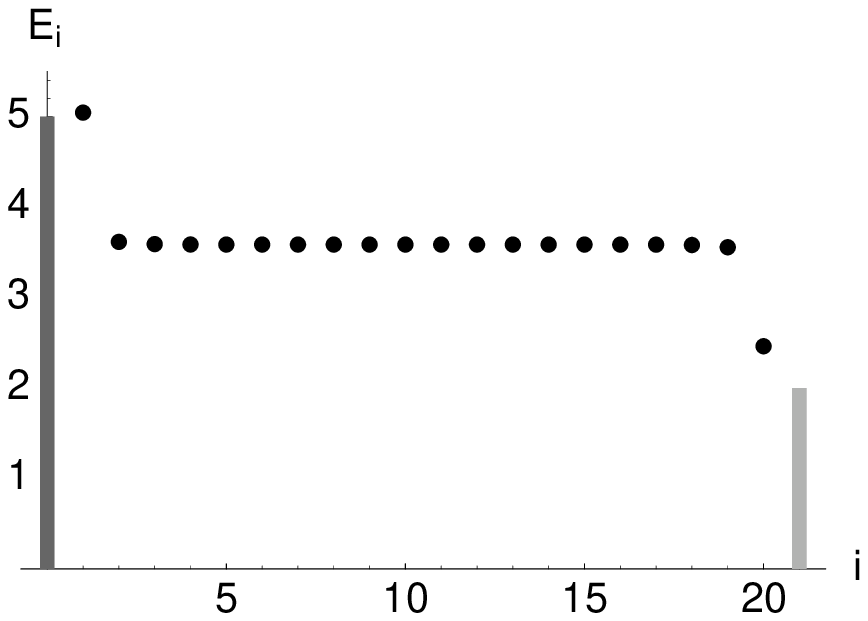}}
\hfill
\subfigure[]{\includegraphics[width=\wzwei]{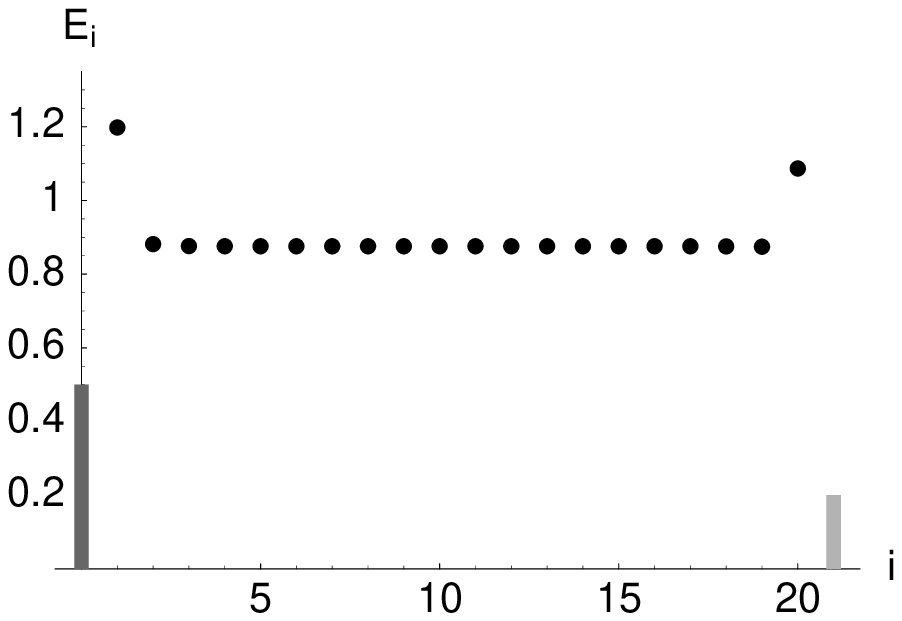}}}
\caption{Energy per site (dots). The bars represent the thermal energies of the respective heat baths. Parameters: $f = 1$, $\gamma = 2$. (a) hight temperatures $T_a=5$, $T_b=2$ (b) low temperatures $T_a=0.5$, $T_b=0.2$}\label{abbTempProfilTiefTemp}
\end{figure}

We transform the correlations of the normal coordinates back to real space and calculate the energy per site, splitting each spring energy to the neighboring sites
\begin{align}\label{eqEn}
E_n &= \frac{\mean{P_n^2}}{2M}+\frac{1}{2}\left(M\omega_n^2+f_{n-1}+f_n\right)\mean{X_n^2}\nonumber\\
&\qquad -f_{n-1}\mean{X_nX_{n-1}}-f_{n}\mean{X_nX_{n+1}}\ .
\end{align}
Numerical results are shown in \autoref{abbTempProfilTiefTemp}.
The energies of the first and the last oscillator are close to the thermal energies of the respective heat bath, and like in the classical investigations (e.~g.\ \cite{clLepriLivi03}), the temperature gradient vanishes inside  the chain. With different coupling parameters $f$ and $\gamma$ one can change the behavior only very close to the boundaries.
In the low temperature case the energies per site are dominated by the zero--point energies. 
The energies of the boundary oscillators are elevated because their effective frequencies are increased by the coupling to the heat baths.

We want to eliminate the zero--point energies and construct a temperature for each lattice site. Like in the previous \subsubsectionname\ we determine effective frequencies with a ground state calculation, using
${E}_0 = \hbar \oeff/2$. Then we assign a temperature $T_R$ using ${E}=\frac{1}{2}\hbar\oeff\coth\left(\frac{\hbar\oeff}{2 k_B T_R}\right)$.

\begin{figure}[bhtp]
{\subfigure[]
{\includegraphics[width=\wzwei]{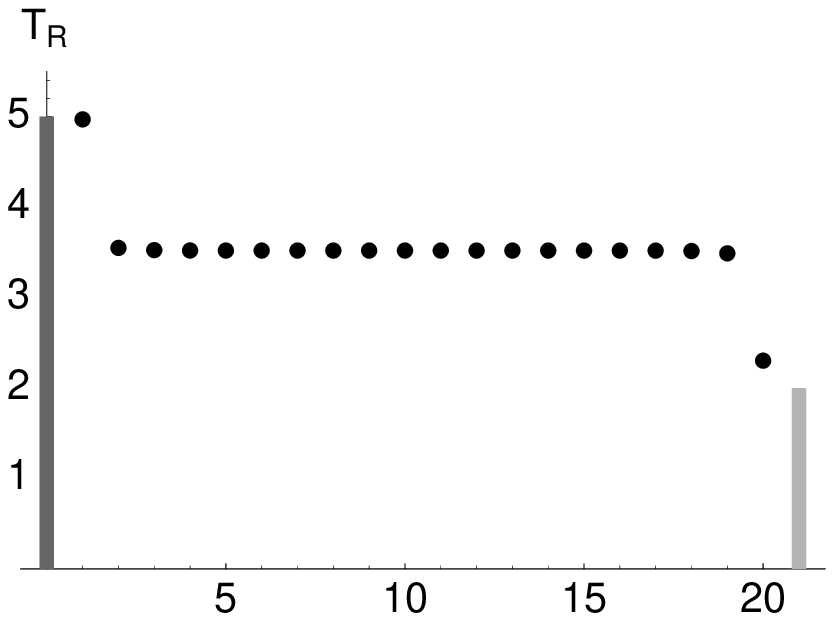}}\hfill
\subfigure[]
{\includegraphics[width=\wzwei]{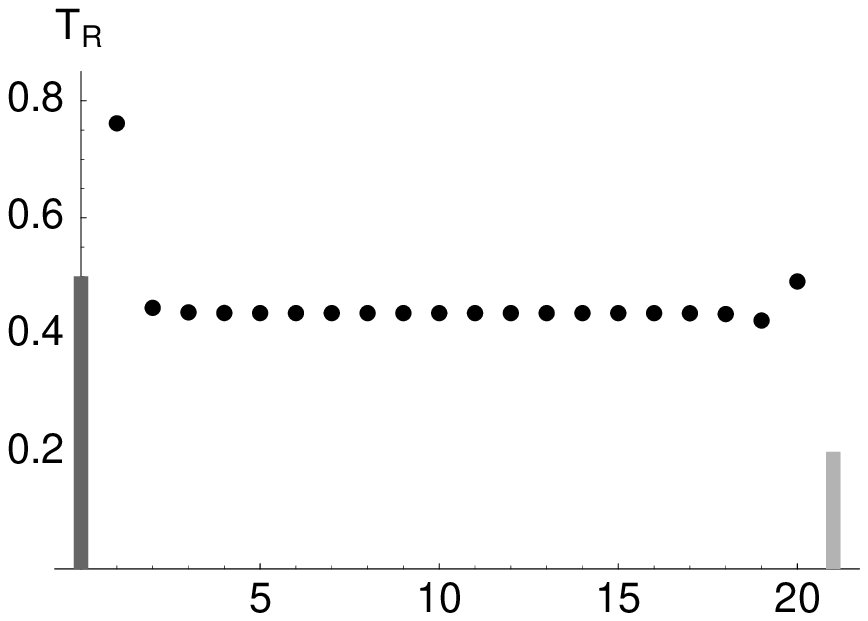}}}
\caption{The temperatures $T_R$ reconstructed from the energies per site and the zero--point energies per site. The bars represent the temperatures of the respective heat baths. Parameters: $f=1$, $\gamma=2$. (a) high temperatures $T_a=5$, $T_b=2$ (b) low temperatures $T_a=0.5$, $T_b=0.2$} \label{abbTRProfilTiefTemp}
\end{figure}
In \autoref{abbTRProfilTiefTemp} the reconstructed temperature is shown.
In the high temperature case (a) the result looks rather the same as in \autoref{abbTempProfilTiefTemp}, because the zero--point energies are not very relevant at high temperatures. In the low temperature case (b) however, the zero--point energies have been compensated and one can see a temperature profile which lies between the two bath temperatures, except for the boundary oscillators. 
Again the temperature gradient is only an exponentially small boundary effect. Note that the interior temperature is again closer to the temperature of the warm heat bath, indicating a higher thermal conductivity at high temperatures.

\subsubsection{The heat flux}\label{ssHeatFlux}
We set up an equation of energy continuity 
\begin{align}
\frac{d}{dt}E_n = J_{n-1,n}-J_{n,n+1}
\end{align}
by differentiating \eqref{eqEn}. In the stationary case we find for the energy fluxes from site $n$ to $n+1$
\begin{equation}
J^\infty_{n,n+1}=\frac{f_{n}}{M}\mean{X_n P_{n+1}}.
\end{equation}
This result agrees with the classical formula \lq power = force $\times$ velocity\rq\ using \lq $\text{force} = f_n(X_{n+1}-X_n)$\rq\ and $\mean{X_n P_n}\propto \frac{d}{d t}\mean{X_n^2}=0$.

\begin{figure}[bhtp]
\includegraphics[width=0.9\linewidth]{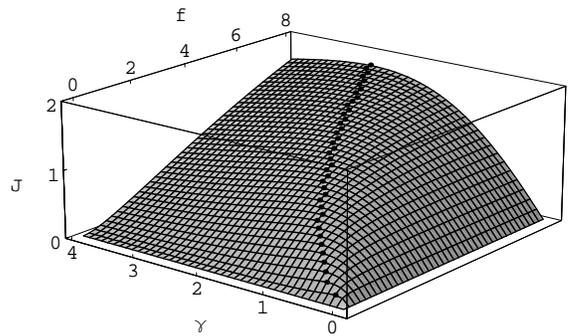}
\caption{The heat flux $J$ as a function of the coupling to the heat baths $\gamma$ and the coupling within the chain $f$. The dots mark the maxima with respect to $\gamma$. $T_a=5$, $T_b=2$}\label{AbbfgammaJ}
\end{figure}

Due to energy conservation $J^\infty_{n,n+1}$ must be independent of $n$, therefore we write $J:=J^\infty_{n,n+1}$.
The heat flux $J$ is one scalar quantity and therefore easier to analyze than the temperature gradient.
\paragraph{Heat flux as a function of the coupling constants}
We calculate the heat flux for different coupling constants $f$ and $\gamma$ and plot the heat flux over the $f$--$\gamma$--plane, see \autoref{AbbfgammaJ}.
In general the heat flux grows with $f$ and $\gamma$, but $f$ and $\gamma$ must match each other:
For a given $f$ the heat flux increases linearly with $\gamma$, passes a maximum at $\gamma_{\text{max}}$ and then vanishes like $\gamma^{-1}$.
This agrees with the behavior found by \citet{Rieder1967} in a classical model without onsite potentials.
In their results $\gamma_{\text{max}}$, the value of $\gamma$ where $J$ is maximal, increases linearly with $f$. In our model $\gamma_{\text{max}}$ starts approximately linearly with $f$ but falls behind for larger $f$.

\begin{figure}[hbtp]
{\subfigure[]
{\includegraphics[width=\wzwei]{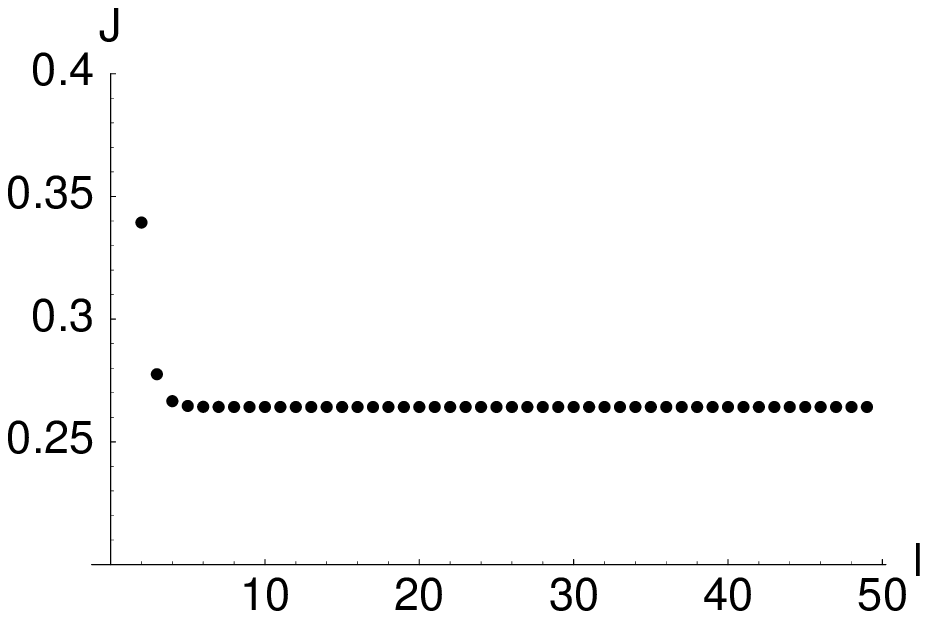}}
\hfill
\subfigure[]
{\includegraphics[width=\wzwei]{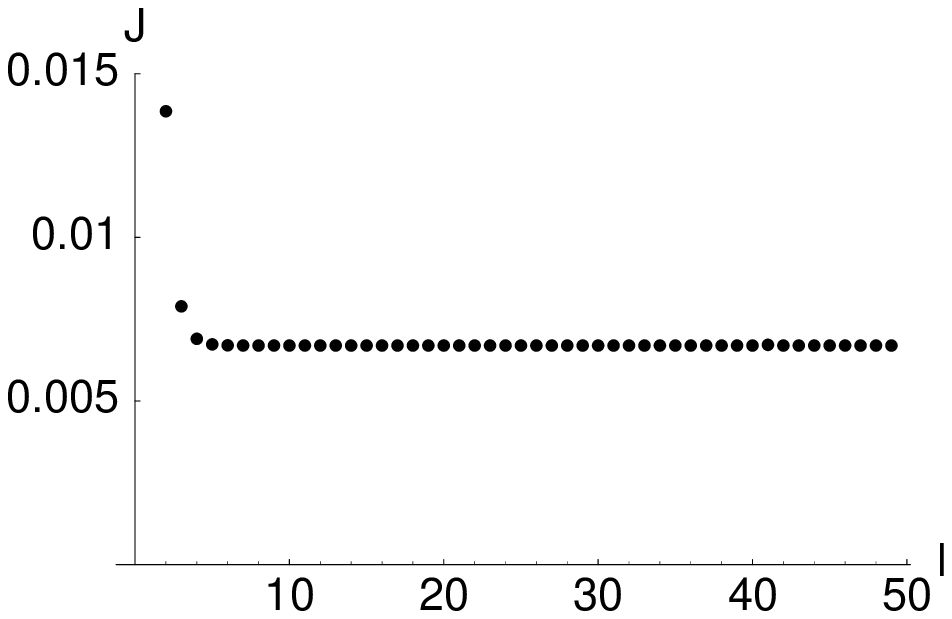}}}
\caption{The heat flux as a function of the chain length $l$. Parameters: $\gamma=2$, $f=1$. (a) high temperatures $T_a=5$, $T_b=2$ (b) low temperatures $T_a=0.5$, $T_b=0.2$}
\label{figOrderJl}
\end{figure}

\paragraph{Heat flux and chain length}\label{ssAllgKettenLg}
In the case of normal heat conduction the heat flux is expected to decrease reciprocal with the chain length at fixed temperature difference $T_a-T_b$. In agreement with the vanishing temperature gradient inside the chain (\ssSecRef{ssTempProfiles}), we find that the heat flux does not decrease with the chain length for $l\gtrsim 5$, \autoref{figOrderJl}. The total heat conductivity rather than the specific conductivity is a constant in the ordered harmonic chain.

\paragraph{Thermal conductivity as a function of temperature}\label{ssStatEnergiestrom}
We are especially interested in the low temperature regime. Therfore we vary the mean temperature with a fixed relative temperature difference $\epsilon=(T_a-T_b)/(T_a+T_b)$ and calculate the conductivity $G\tt{th}:={J}/(T_a-T_b)$ for each temperature. 

\begin{figure}[hbtp]
\centering
\subfigure{\includegraphics[width=.8\linewidth]{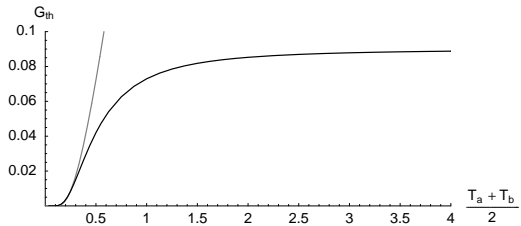}}\hfill
\caption{Black: The thermal conductivity as a function of temperature. Gray: $c \left[\exp\left(\hbar\omega_0/\kT\right) -1 \right]^{-1}$.
Parameters: $\epsilon=0.01$, $f=1$, $\gamma=2$}\label{abb:varTm}
\end{figure}

Results are shown in \autoref{abb:varTm}. In the high temperature regime the conductivity is a constant like in the classical case. In the low temperature regime it breaks down and behaves similarly to the Bose--Einstein occupation numbers of the lowest normal frequencies (gray line). In this case the degrees of freedom of the chain are simply frozen out.

This behavior is typical for optical phonons. In our model the low normal frequencies start with the value $\omega_0$, which means all normal modes are frozen if $k_B T \lesssim \hbar \omega_0$. If we set $\omega_0$ to zero in the inner chain, the normal frequencies start at zero and we find the temperature dependence of acoustic phonons $\Gth \propto T^3$, which was also found in \cite{Talkner1990_2}.

\subsection{Disordered Chains}\label{ssResultsDisorder}
It is well known from classical works that disorder can lead to a finite temperature gradient inside the chain, e.~g.\ \cite{Verheggen1979,disoDhar01, clLepriLivi03}.
There are different possibilities to bring disorder into play:
\begin{itemize}
\item A common choice in the literature is to choose the masses of the oscillators randomly. This corresponds to isotopical disorder in nature. 
\item Random frequencies of the onsite potentials $\omega_i$
\item Random coupling constants $f_i$ between the oscillators of the chain.
\end{itemize}
To simplify matters we confine ourselves to disorder in the couplings $f_i$ in this article. The $f_i$ are chosen from a Gaussian distribution with mean $\bar f$ and width $\sigma_f$, and a cutoff that guarantees that the $f_i$ are always positive. In the following we use the notation $f=\bar f \pm \sigma_f$.
Temperature profile, heat flux etc.\ are calculated for many realizations of disorder and averaged over. In the following $\mean{\cdot}$ denotes the ensemble average.

As indicated in \sSecRef{ssSymm}, the symmetry of the chain is relevant. Therefore we always compare the results of the symmetrical and the unsymmetrical disordered chain in this \subsectionautorefname.

\subsubsection{Normal modes in disordered chains}
\begin{figure}[hbtp]
\includegraphics[width=0.8\linewidth]{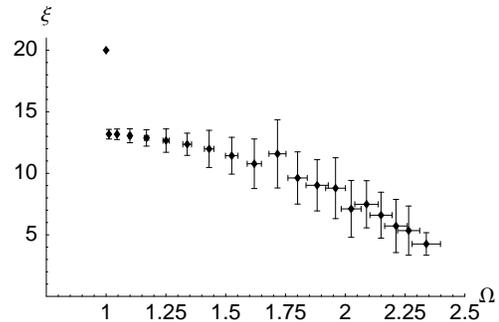}
\caption{The localization length $\xi$ in dependence of normal frequencies $\Omega$ for an unsymmetrical chain with $f_i=1 \pm 0.2$ averaged over 50 realizations, the error bars show the standard deviation.}\label{abbLoc}
\end{figure}

The disorder changes the normal modes from standing waves to localized states. As we are considering disorder in the couplings $f_i$, the high frequency modes are affected the most by disorder. We calculate the localization length $\xi$ in units of the lattice constant as the inverse of the participation number $p$:
\begin{align}
\xi_i := p_i^{-1} := \bigg[\sum_{j=1}^l Y_{ij}^4 \biggr]^{-1}
\end{align}
The center--of--mass mode, which is extended over the whole chain yields $\xi = l$, a fully localized mode yields $\xi=1$.
Results for an unsymmetrical disordered chain are shown in \autoref{abbLoc}. The center--of--mass mode has $\xi=l$, all the other modes have smaller localization lengths, which decrease with frequency. 

In the symmetric chain the normal modes are constrained to be symmetric or antisymmetric, and are therefore less localized.

\subsubsection{Occupation numbers}
\begin{figure}[hbtp]
\subf a {\includegraphics[width=0.74\linewidth]{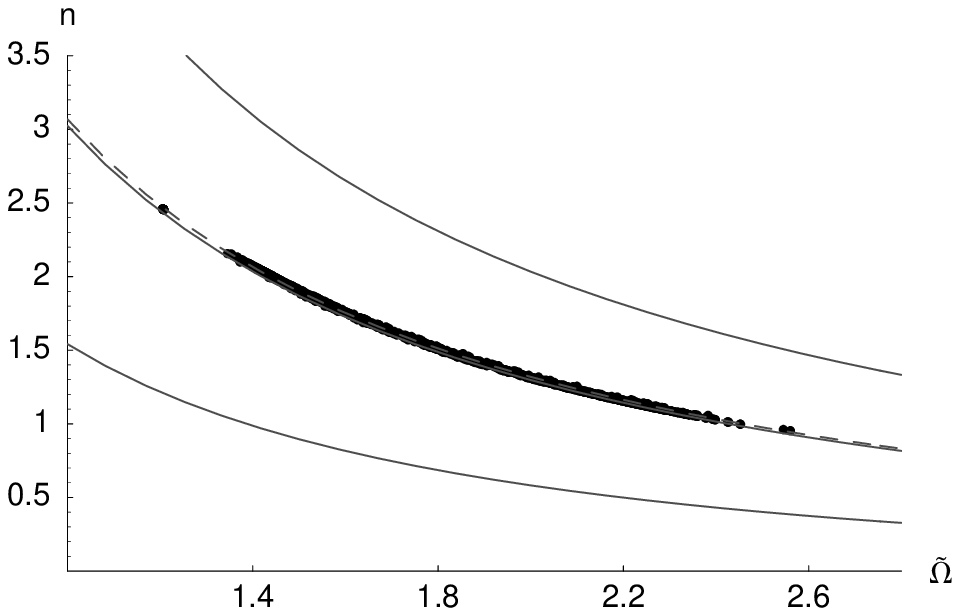}}\\
\subf b {\includegraphics[width=0.74\linewidth]{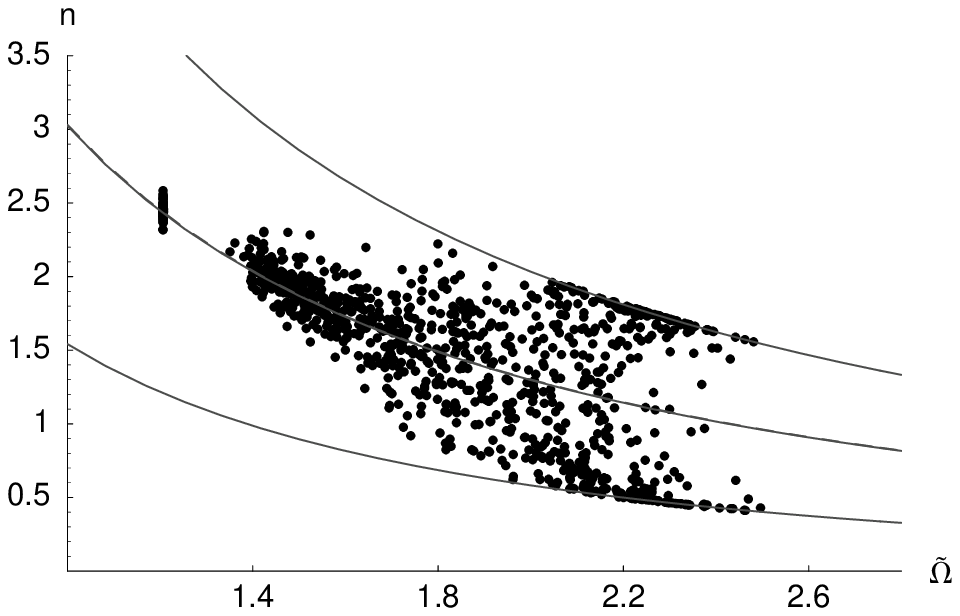}}\\
\subf c {\includegraphics[width=0.74\linewidth]{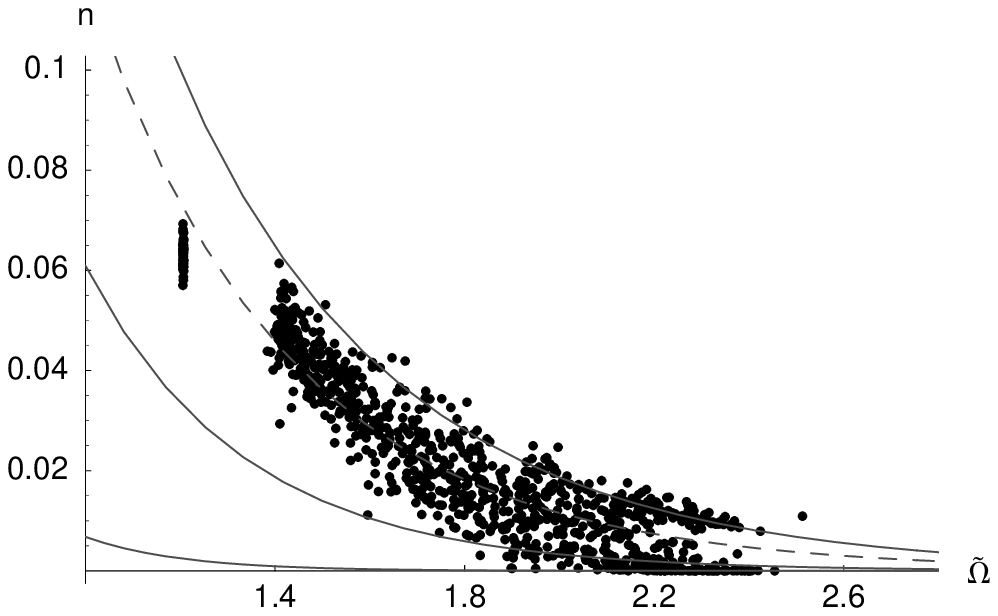}} 
\caption{Occupation numbers from 50 realizations of disorder. The lines show Bose--Einstein distributions corresponding to the mean temperature $(T_a+T_b)/2$, to the bath temperatures $T_a$ and $T_b$ and to the average temperature of the normal modes (dashed).
Parameters: $\omega_0 = 1$, $\gamma=2$, $T_a = 5$, $T_b= 2$, $f=1\pm 0.2$.
(a) symmetrical chain, $T_a=5$, $T_b=2$
(b) unsymmetrical chain, $T_a=5$, $T_b=2$
(c) unsymmetrical chain, $T_a=0.5$, $T_b=0.2$}\label{abbSymmBrBes}
\end{figure}

In \autoref{abbSymmBrBes} the occupation numbers of disordered chains are shown. In the symmetrical chain the frequencies are smeared by disorder, compared to the ordered case in \autoref{figOccNumbers}~(a).
At the same time the occupation numbers still agree with the Bose--Einstein distribution at the mean temperature.

In the unsymmetrical case, \autoref{abbSymmBrBes}~(b), this changes: The distribution broadens and the points lie between the Bose--Einstein distributions corresponding to $T_a$ and $T_b$. It is striking that the points belonging to the high frequency modes tend to lie close to either bath temperature.
This is due to the fact that the strongly localized high--frequency modes are not restricted to be symmetric or antisymmetric any more. Therefore they are coupled much more strongly to the bath they lie closer to.

\begin{figure}[hbtp]
\subfigure[]
{\includegraphics[width=\wzwei]{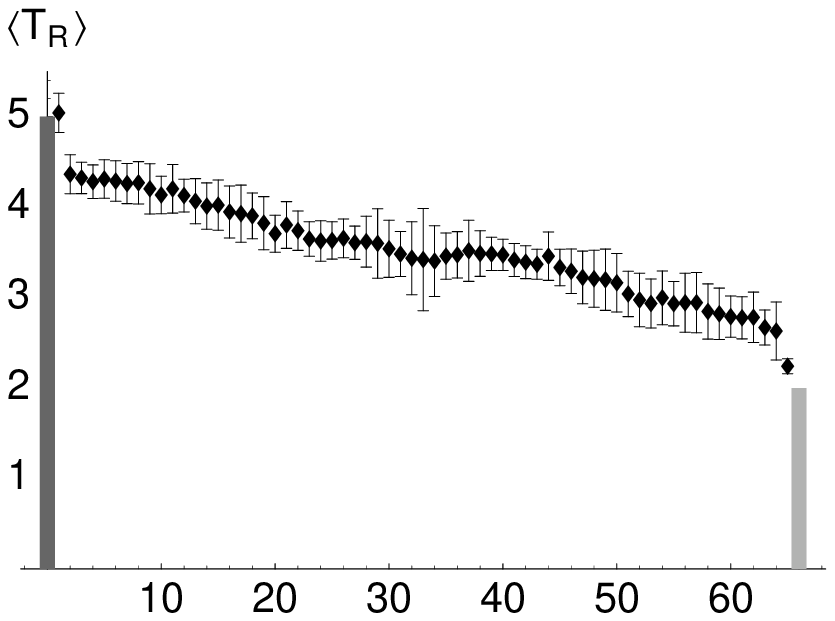}}\hfill
\subfigure[]
{\includegraphics[width=\wzwei]{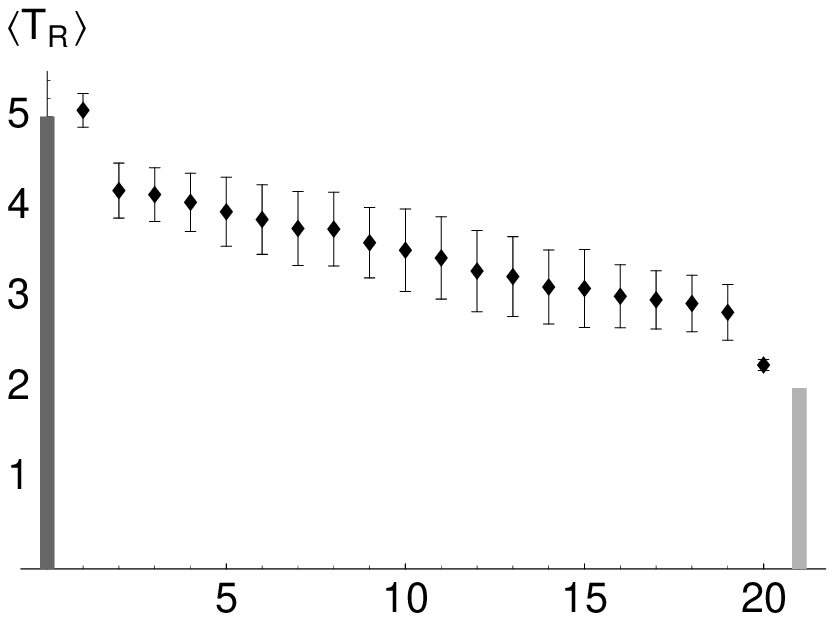}}
\caption{Averaged temperature profile. The error bars show the standard deviation of the temperature at each lattice site. Parameters: $f=1\pm 0.2$, $\omega_0 = 1$, $\gamma=2$, $T_a = 5$, $T_b= 2$. (a) symmetrical, $l=65$, 22~realizations
(b) unsymmetrical, $l=20$, 50~realizations}\label{abbSymmBrTempAvg}
\end{figure}

\subsubsection{Temperature profile and heat flux}
\begin{figure}[hbtp]
{\includegraphics[width=0.7\linewidth]{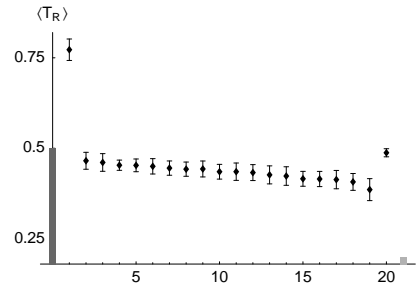}}
\caption{Temperature profile in the unsymmetrical case averaged over 50 realizations. In the symmetrical case the temperature gradient is about half as steep. Parameters: $f=1\pm 0.2$, $l=20$, $\omega_0$~=~$1$, $\gamma=2$, $T_a = 0.5$, $T_b= 0.2$}\label{abbSymmBrTempLowTAvg}
\end{figure}

Again we transform to real space and calculate the energy distribution.
There are great differences between the single realizations of disorder, so one has to average over many realizations in order to obtain comparable results, see  \autoref{abbSymmBrTempAvg}. The temperature gradient is enhanced in the disordered chain and stays finite, even for long chains. In the unsymmetrical case the distribution is wider and the average temperature gradient is steeper.
In the low temperature disordered case, \autoref{abbSymmBrTempLowTAvg}, the contact resistance at the cold bath is large and the temperature gradient within the chain remains small.

\begin{figure}[hbtp]
\subfigure[]{\includegraphics[width=\wzwei]{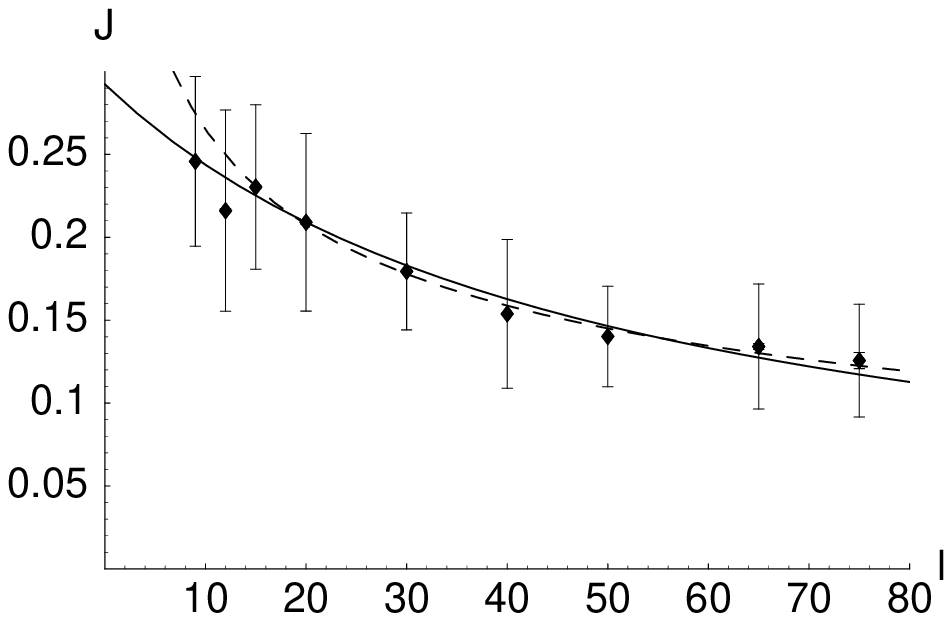}}\hfill
\subfigure[]
{\includegraphics[width=\wzwei]{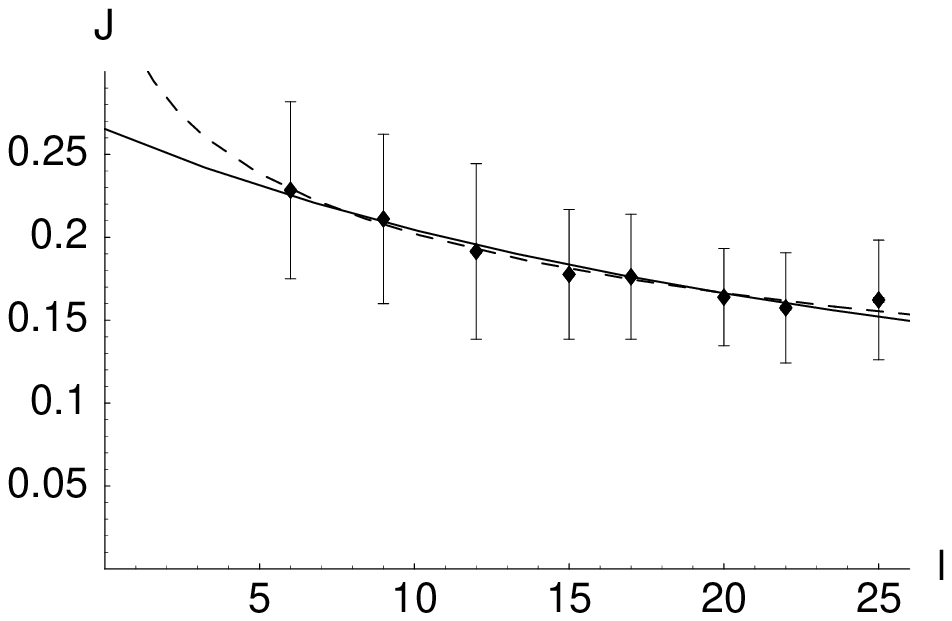}}
\caption{The heat current as a function of the chain length $l$. Each point shows the average over $k$ realizations, where $22 \leq k \leq 120$. The large errorbars show the width of the heat flux distribution $\sigma_J$ and the small errorbars show the error of the average $\sigma_J/(k-1)$. The solid line is a fit for normal bulk resistance proportional to $l$ and the dashed line is a fit for a bulk heat resistance proportional to $\sqrt{l}$. In the symmetrical case (a), the chain lengths are large enough to indicate that the data fits better to the bulk heat resistance proportional to $\sqrt{l}$. In the unsymmetrical case (b) the achieved chain lengths are insufficient.
Parameters: $\omega_0 = 1$, $\gamma=2$, $T_a = 5$, $T_b= 2$, $f=1\pm0.2$.
}\label{figDisorderJl}
\end{figure}
The heat flux is reduced in the disordered case, compare \autoref{figDisorderJl} and \autoref{figOrderJl}~(a). The main difference compared to the ordered case (see \autoref{figOrderJl}) is, that the heat flux is not independent of the chain length for $l \gtrsim 5$ any more. In the symmetrical case it was possible to calculate the correlations for a chain length up to $l=75$ in a reasonable time. 
One can try to determine the asymptotic behavior from this data. For heat conduction according to Fourier's law we would expect a heat flux proportional to $(R_c + R_{\text{bulk}} l)^{-1}$ with a contact resistance $R_c$ and the resistance of the inner chain $R_{\text{bulk}} l$. 
For harmonic chains the classical asymptotic resistance in proportional to $\sqrt{l}$ \cite{Verheggen1979,disoDhar01}. The dashed lines are fits to this behavior. 
Due to the wide distribution of the heat currents and the finite lengths of the considered chains, it is not evident which fit describes the asymptotic behavior, but at least in the case of symmetric chains our model seems to show $\sqrt{l}$ asymptotic behavior (dashed line).

\subsection{Entanglement}\label{ssEntanglement}
Beside the zero point energies and the Bose--Einstein statistics of the occupation numbers, entanglement is another relevant quantum mechanical feature that can be observed in our model.

We use the logarithmic negativity, which is a measure for the entanglement between two parts of a system, which can be calculated from the correlations between the coordinates and momenta of the system \cite{entPlenioHartleyEisert04,entVidalWerner02}.

We set up the covariance matrix containing all the correlations between coordinates and momenta in real space. Then the system is divided into two parts $\A$ and $\B$ and the covariance matrix is partially transposed with respect to $\B$. The logarithmic negativity is then given by
\begin{equation}
N=-\sum_j \log_2(\min(1,|\gamma_j|))\ ,
\end{equation}
where the $\gamma_j$ are the symplectic eigenvalues of the partially transposed covariance matrix.
We use the notation
$N_k$ for the logarithmic negativity of the subsystems $\A = \menge{X_1,\ldots,X_k}$ and $\B=\menge{X_{k+1},\ldots,X_l}$.
Other divisions, e.~g.\ taking every second oscillator or performing the division in normal mode space, are also possible, but their physical meaning is not obvious.

\subsubsection{Entanglement in dependence of the couplings $f$ and $\gamma$}
\begin{figure}[hbtp]
\includegraphics[width=0.73\linewidth]{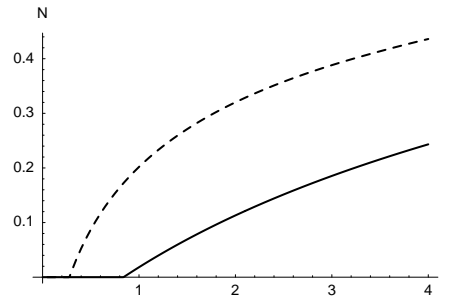}
\caption{Stationary logarithmic negativity as a function of the coupling $f$ inside the ordered chain.
Solid line: $N_1$, dashed line: $N_5$.
Parameters: $ l=20$, $\gamma=2$.}\label{figNf}
\end{figure}

We start with the ordered case.
Figure \ref{figNf} shows the entanglement for different divisions of a chain of length 20 as a function of the coupling within the chain $f$.
With $f=0$ the oscillators are not coupled at all and there is no entanglement. Increasing $f$ favors entanglement. For each $N_k$ there is a threshold coupling where entanglement starts.
In $N_1$ and $N_{l-1}$ one of the subsystems is a single oscillator at the end of the chain coupled directly to a bath. In this case we observe a lower logarithmic negativity and a higher threshold coupling for the onset of entanglement. All the other $N_k$ with $1<k<l-1$ behave very similarly.

\subsubsection{Entanglement as a function of temperature}\label{ssEntDisorderTemperature}
We expect the logarithmic negativity to decrease and finally vanish with increasing temperature. Additionally to the ordered case we want to investigate the entanglement in the unsymmetrical disordered case. Therefore we vary the mean temperature $T_m=(T_a+T_b)/2$ for each realization of disorder.
Another quantity marking the transition from the quantum mechanical regime is the heat conductivity (see \ssSecRef{ssHeatFlux}) which will be observed simultaneously. 

\begin{figure}[hbtp]
\subf a
{\includegraphics[width=0.7\linewidth]{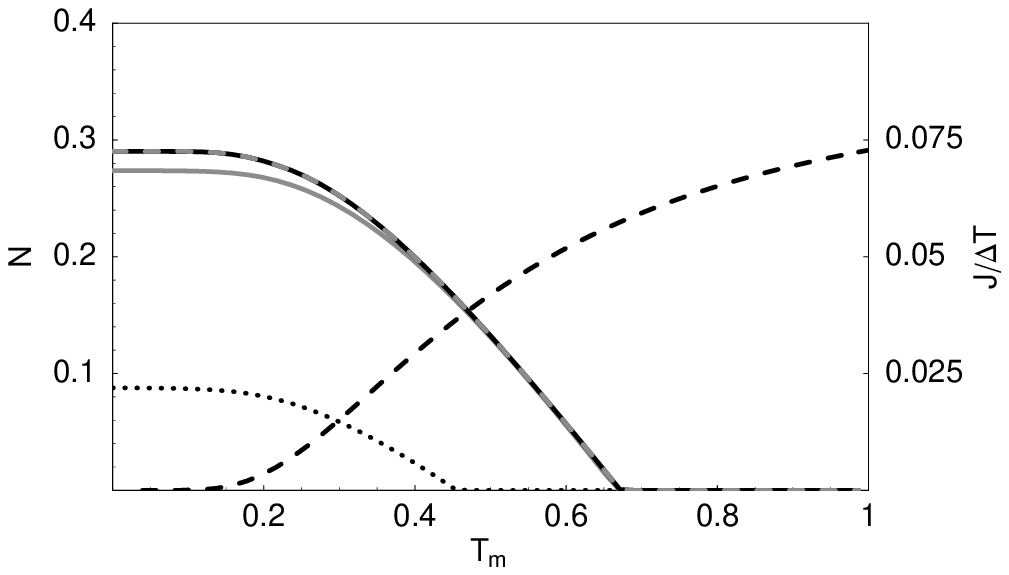}}\\
\subf b
{\includegraphics[width=0.7\linewidth]{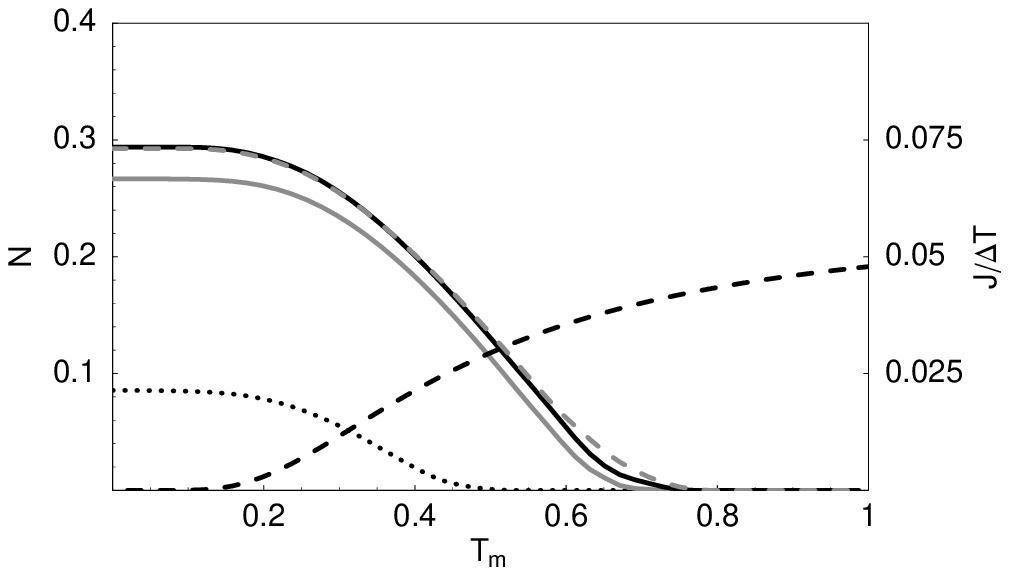}}\\
\subf c	
{\includegraphics[width=0.7\linewidth]{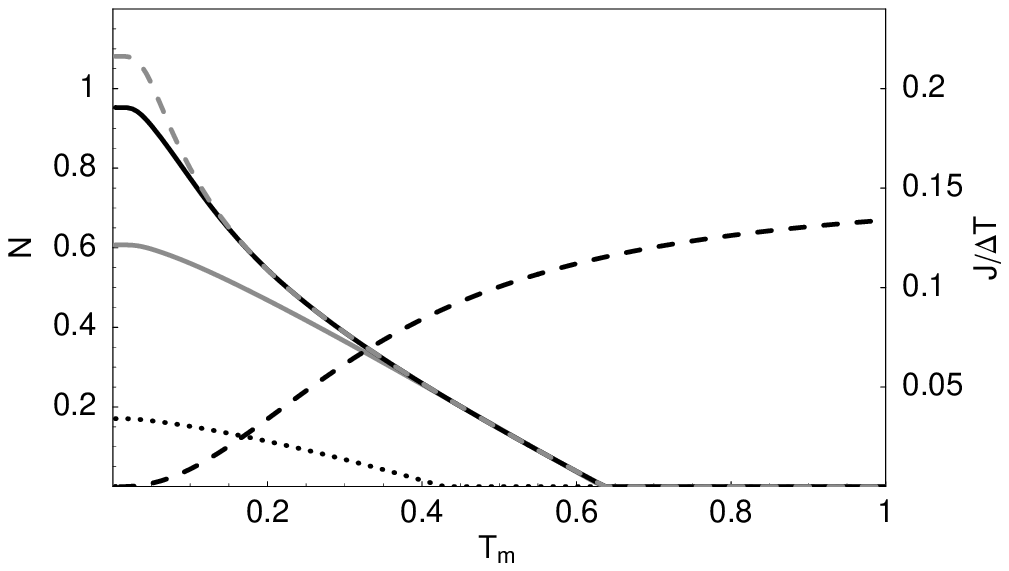}}
\caption{The heat conductivity $J/\Delta T$ (dashed line) and some entanglements as a function of temperature. Dotted line: $N_1$; gray line: $N_2$; solid line: $N_{5}$; gray dashed line: $N_{10}$. Parameters from \eqref{eqTVarParameter}. 
(a) Ordered Chain, $\epsilon=0.1$
(b) Unsymmetrical disordered chain, $f=1\pm 0.2$, averaged over 20 realizations, $\epsilon=0.1$
(c) Ordered Chain, no onsite potential within the chain, $\epsilon=0.01$
}\label{abbEntGTemp}
\end{figure}

$T_a$ and $T_b$ are chosen as $(1\pm\epsilon)T_m$ with $\epsilon=0.1$. The further parameters are chosen
\begin{align}\label{eqTVarParameter}
l&=20 &  \Gamma &=10 & \gamma &=2 & \bar f&=1 & \sigma_f &=0.2 & \omega_0&=1 \ .
\end{align}
In \autoref{abbEntGTemp}~(a) the conductivity and entanglements for different divisions of the chain are plotted for an ordered chain. The entanglements with one subsystem consisting of a single oscillator $N_1$ and $N_{19}$ have lower values than the others, as observed already in the previous \subsubsectionname. All $N_k$ are initially constant and start decreasing at $T_m\approx 0.2$. $N_1$ and $N_{19}$ reach zero at $T_m\approx 0.45$, the others follow at $T_m\approx 0.67$. In the same temperature range the heat conductivity $J/\Delta T$ rises.

In the disordered case (\autoref{abbEntGTemp}~(b)) there are only minor changes to the entanglement in each realization of disorder. The sharp transition to zero is washed out by the average. As seen before, the heat conductivity $J/\Delta T$ is lower in the disordered case, but it shows essentially the same temperature dependence.

The plateau we observe at low temperatures results from the frequencies of the normal modes starting from $\omega_0$. In the case of a chain with onsite potentials only at the ends of the chain, \autoref{abbEntGTemp}~(c), the logarithmic negativity reaches higher values at low temperatures, but their decrease starts already at lower temperatures.

\subsection{Time evolution}\label{ssTime}
\begin{figure}[hbtp]
\includegraphics[width=.8\linewidth]{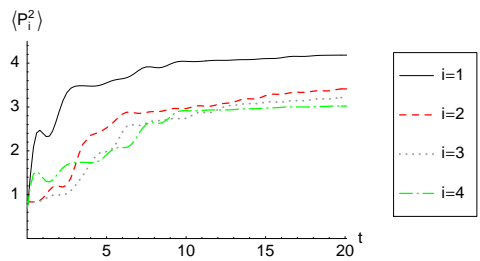}
\caption{(Color online) The time evolution of the diagonal momentum correlations. Parameters: $f=1$, $\gamma=0.5$, $T_a=5$, $T_b=2$}\label{figPPt}
\end{figure}

Finally we shortly want to study some time dependencies of the system. For simplicity we inspect a short ordered chain, consisting of only four oscillators. We evaluate the time dependent correlations (like equation \eqref{eqAllgYYKorrelation}) and study e.~g.\ the diagonal momentum correlations, which are proportional to the kinetic energy, see \autoref{figPPt}.
Under some oscillations each lattice site gains energy until it reaches the value corresponding to the stationary temperature profile.

\begin{figure}[hbtp]
\includegraphics[width=.8\linewidth]{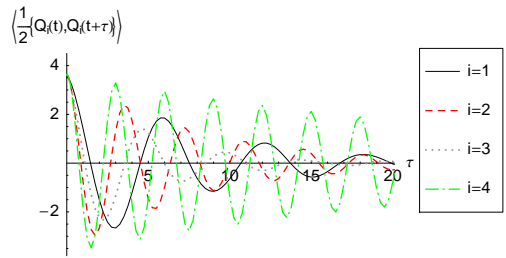}
\caption{(Color online) The time shifted diagonal momentum correlations in normal coordinates in the limit $t\rightarrow\infty$. Parameters like in \autoref{figPPt}}\label{figQQtau}
\end{figure}

Furthermore we generalize equation \eqref{eqAllgYYKorrelation} for correlations of two coordinates or momenta at different times. In \autoref{figQQtau} results are shown for the time shifted autocorrelation functions of the momenta in normal coordinates in the stationary limit. Each correlation performs damped oscillations with its slightly detuned frequency. Each normal coordinate loses energy by damping which is replaced by incoherent noise. The stronger the influence of damping and noise to a normal coordinate, 
the faster it loses memory of its history and the autocorrelation function decays.
Normal modes with large amplitudes at the ends of the chain experience the strongest damping.

\section{Summary}
We have treated disordered harmonic chains with the Quantum Langevin formalism in the strong coupling regime.
The strong coupling to the heat baths led to a renormalization of the normal frequencies.

In ordered chains the occupation numbers calculated with these frequencies approximately follow the Bose--Einstein statistics, even in the strong--coupling non--equilibrium regime.
The heat flux follows from non--diagonal correlations between coordinates and momenta.

Symmetric disorder does not change the occupation numbers qualitatively, because the normal coordinates are constrained to have the same amplitude at both ends of the chain.
With breaking the left--right symmetry this changes. The localization of most of the modes enhances the effect of strongly asymmetric coupling to the heat baths.

In the low temperature regime the energy distribution in the chain is dominated by zero--point energies that have to be taken into account for constructing the local temperature $T_R$.

In the limit of ordered chains we have recovered the vanishing temperature gradient and the length--independent heat flux known from classical models. 
In disordered chains the temperature gradient is finite and the heat flux decreases with the length. This decrease seems to be slower than $l^{-1}$, following the classical prediction for harmonic disordered chains $l^{-\frac{1}{2}}$, but the asymptotic behavior could not be identified clearly due to numerical restrictions.

Characteristic quantum mechanical features are the freezing of the heat conductivity, that behaves typical for optical phonons in our model. In the same temperature range where the heat conductivity freezes, entanglement appears.

\appendix

\section{The noise response functions in the symmetrical case}\label{sRauschAntwort}
\subsection{The structure of the response functions}
\def\Om#1{(s^2+\Omega_{#1}^2)}
\def\Ds{\frac{s \hat \gamma(s)}{M}}
\def\Gm#1#2{2 G_{1#1}G_{1#2}\Ds}
\def\Divisor#1{\hat D_{\p{#1}}(s)}
We want to calculate only the noise response functions and therefore omit the initial conditions in \eqref{eqLaplaceYi}. We use the symmetry relations \eqref{eqSymmG} and get
\begin{equation}\label{eqAppendix1}
\Om{i} \hat Y_i(s) + \sum_{j\in\sf{i}}\Gm i j\hat Y_j(s)=\frac{G_{1i}}{M}\hat\eta_{\p{i}}(s) \, .
\end{equation}
Dividing this equation by $G_{1i}$ and subtracting the same equation with $i$ replaced by $n$ eliminates the $\hat \gamma$ and the $\hat\eta$ terms and yields
\begin{equation*}
\frac{\Om n}{G_{1n}}\hat Y_n(s) = \frac{\Om i}{G_{1i}}\hat Y_i(s) .
\end{equation*}
Using this equation for eliminating all $\hat Y_j$ with $j\neq i$ from \eqref{eqAppendix1} yields
\begin{equation}
\hat Y_i(s)=\frac{1}{M}\hat F_{i}(s)\hat\eta_{\p{i}}(s) \, ,
\end{equation}
with
\begin{equation*}
\hat F_{i}(s)=\frac{G_{1i}}{\Om i}\left[1+\sum_{j\in\sf{i}}\frac{2 G_{1j}^2}{\Om j}\Ds \right]^{-1} .
\end{equation*}
Extracting the common denominator $\Divisor{i}$, which is equal for all response function from the same symmetry, yields the following form
\begin{equation}
\hat F_{i}(s)={(s+\Gamma)G_{1i}\prod_{\stackrel{j\in\sf{i}}{j\neq i}}\Om j }\Big/{\Divisor{i}} \, .
\end{equation}
The numerator contains any $\Om i$ from the same symmetry, except for the term with its own frequency. It is proportional to the amplitude of the mode at the end of the chain $G_{1i}$.

In the unsymmetrical case this analytic calculation of the noise functions is not possible due to mixing of the symmetry families. Therefore Cramer's rule has to be applied explicitly, which requires a higher numerical effort.

\subsection{The roots of the denominator}
The denominator of the response functions reads

\begin{align}\label{eqApp2}
\Divisor{i}&=(s+\Gamma)\prod_{j\in\sf{i}}\Om{j} \nonumber \\&\quad
+2s\frac{\gamma\Gamma}{M}\sum_{j\in\sf{i}}G_{1j}^2\prod_{k\neq j}\Om{k} .
\end{align}

We will prove that the real parts of the roots of $\Divisor{i}$ are negative.
As $s=\pm i\Omega_j$ and $s=0$ are no roots, we can search for the roots of $\Divisor{i}/ \left[ \prod_{j\in\sf{i}}\Om{j}\Gamma s\right]$, which implies
\begin{align}\label{eqCharNullst}
\hat u(s):=\frac{1}{\Gamma}+\frac{2\gamma}{M}\sum_{k\in\sf(i)} \frac{G_{1k}^2}{\Om{k}} = -\frac{1}{s}
\end{align}
Without loss of generality we assume that $\Im(s)>0$. Now we assume $\Re(s)>0$, which implies $\arg(s) = \arg(-1/s)\in(0,\pi/2)$.
On the other hand we find $\arg(s^2+\Omega_j^2)\in (0,\pi)$ $\Rightarrow$ $\arg(1/\Om j)\in (-\pi,0)$ $\Rightarrow$ $\arg(\hat u(s))\in(-\pi,0)$. Therefore \eqref{eqCharNullst} cannot be fulfilled and the assumption $\Re(s)>0$ is disproven.

\bibliography{/home/btp3/btp304/Literatur/bibtex/literaturQHC}

\end{document}